\documentclass[reprint,twocolumn,showpacs,preprintnumbers,superscriptaddress,amsmath,amssymb]{revtex4-2}
\usepackage{graphicx}
\usepackage{siunitx}
\usepackage{xcolor}
\usepackage{amsmath,amssymb}
\usepackage[T1]{fontenc}
\usepackage{lmodern}
\usepackage[utf8]{inputenc}
\usepackage[english]{babel}
\usepackage{natbib}
\usepackage{hyperref}
\hypersetup{colorlinks=true,citecolor={blue},linkcolor={blue},urlcolor={blue}}
\usepackage{siunitx}

\begin{document}

\title{Tailoring Flatband Dispersion in Bilayer Moiré Photonic Crystals}

\author{Chirine Saadi}
\affiliation{Ecole Centrale de Lyon, INSA Lyon, Universit\'e  Claude Bernard Lyon 1, CPE Lyon, CNRS, INL, UMR5270, Ecully 69130, France}

\author{ Sébastien Cueff}
\affiliation{Ecole Centrale de Lyon, INSA Lyon, Universit\'e  Claude Bernard Lyon 1, CPE Lyon, CNRS, INL, UMR5270, Ecully 69130, France}

\author{Lydie Ferrier}
\affiliation{Ecole Centrale de Lyon, INSA Lyon, Universit\'e  Claude Bernard Lyon 1, CPE Lyon, CNRS, INL, UMR5270, Ecully 69130, France}

\author{Aziz Benamrouche}
\affiliation{Ecole Centrale de Lyon, INSA Lyon, Universit\'e  Claude Bernard Lyon 1, CPE Lyon, CNRS, INL, UMR5270, Ecully 69130, France}

\author{Maxime Gayrard}
\affiliation{Ecole Centrale de Lyon, INSA Lyon, Universit\'e  Claude Bernard Lyon 1, CPE Lyon, CNRS, INL, UMR5270, Ecully 69130, France}

\author{Emmanuel Drouard}
\affiliation{Ecole Centrale de Lyon, INSA Lyon, Universit\'e  Claude Bernard Lyon 1, CPE Lyon, CNRS, INL, UMR5270, Ecully 69130, France}

\author{ Xavier Letartre}
\email{xavier.letartre@ec-lyon.fr}
\affiliation{Ecole Centrale de Lyon, INSA Lyon, Universit\'e  Claude Bernard Lyon 1, CPE Lyon, CNRS, INL, UMR5270, Ecully 69130, France}

\author{Hai Son Nguyen}
\email{hai-son.nguyen@ec-lyon.fr}
\affiliation{Ecole Centrale de Lyon, INSA Lyon, Universit\'e  Claude Bernard Lyon 1, CPE Lyon, CNRS, INL, UMR5270, Ecully 69130, France}
\affiliation{Institut Universitaire de France (IUF), F-75231 Paris, France}

\author{Ségolène Callard}
\email{segolene.Callard@ec-lyon.fr}
\affiliation{Ecole Centrale de Lyon, INSA Lyon, Universit\'e  Claude Bernard Lyon 1, CPE Lyon, CNRS, INL, UMR5270, Ecully 69130, France}

\begin{abstract}
In this study, we experimentally investigate the photonic dispersion in one-dimensional moiré structures formed by stacking two photonic crystal slabs with slightly different periods, separated by a carefully controlled subwavelength optical spacer. Angle-resolved reflectivity measurements reveal moiré bands arising from the interplay between intra- and inter-layer coupling mechanisms of guided modes mediated by the moiré superlattice corrugation. By precisely adjusting the refractive index contrast through the filling factor of the photonic crystals, we continuously tune intralayer coupling while keeping interlayer coupling constant. Consequently, we experimentally demonstrate the evolution of moiré minibands into flatbands characterized by minimal dispersion bandwidth. All experimental results show good agreement with numerical simulations. Our findings not only confirm theoretical predictions but also provide a practical approach for realizing photonic flatbands in silicon-based moiré superlattices operating in the telecom wavelength range. This work paves the way toward harnessing flatband physics in advanced optoelectronic applications such as lasers and optical sensors.

\end{abstract}

\maketitle

\section{Introduction}
Flatbands are dispersionless energy bands in periodic media, characterized by macroscopic degeneracy and suppressed transport under standard conditions\,\cite{derzhko2015strongly,leykam2018perspective,danieli2024flat}. They typically emerge in finite-range hopping lattices where destructive interference preserves compact localized states and inhibits wavefunction delocalization. These features make flatbands ideal for studying strongly correlated phases and exotic quantum phenomena. Photonic platforms have recently become a promising arena for flatband physics. Early studies focused on Lieb and Kagome photonic lattices based on tight-binding analogs of supra-wavelength photonic “atoms”\,\cite{vicencio2015observation,mukherjee2015observation,Zong2016,xia2016demonstration}. More recently, in subwavelength-scale photonic crystals, “local” flatbands have been realized in specific regions of the Brillouin zone, arising from vertical symmetry breaking\,\cite{nguyen2018symmetry} or lattice distortion that mimics photonic Landau levels\,\cite{Barczyk2024,Barsukova2024}.

To go beyond local flattening and achieve flatbands across the entire Brillouin zone, moiré engineering has emerged as a powerful strategy. This approach draws inspiration from twisted bilayer graphene, where flatbands appear at magic angles due to moiré hybridization gaps that isolate them from Dirac bands, enabling unconventional superconductivity\,\cite{bistritzer2011moire,Tarnopolsky2019,lisi2021observation,cao2018unconventional}, quantum Hall effects\,\cite{Zhang2021}, and topological phases\,\cite{Choi2021}. Theoretical studies predict that similar mechanisms in twisted 2D photonic crystals can lead to flatbands and enhanced confinement\,\cite{Oudich2021,dong2021flat,tang2021modeling,tang2022chip}, with strong interlayer coupling supporting robust, potentially topological, flatbands even at large twist angles\,\cite{Yi2022}. Additional moiré-induced phenomena include chiral emission\,\cite{Lou2021,Zhou2025} and vortex generation\,\cite{Zhang2023}. Experimentally, most work has focused on effective moiré potentials rather than true bilayer superlattices. Initial demonstrations used monolayer architectures with superimposed lattices, enabling lasing\,\cite{article32,article36} and harmonic generation\,\cite{Wang2024}, where flatbands boost confinement and nonlinear effects. However, continuous tuning of dispersion and direct observation of bandwidth collapse at magic angles remain unrealized. Full control over intra- and interlayer coupling requires bilayer structures with true geometric degrees of freedom. Recent efforts have fabricated twisted bilayer photonic crystals, demonstrating twist-angle–tunable dispersion and moiré scattering in microwave\,\cite{Lou2022} and telecom\,\cite{tang2023experimental,Tang2025} regimes. Yet, despite high tunability, flatbands have not yet been experimentally observed.

To simplify the fabrication challenges while preserving essential physical phenomena, an alternative approach is employing one-dimensional (1D) crystals instead of 2D structures. In our previous theoretical work, we showed that periodic mismatches between two stacked 1D photonic crystals can also yield photonic moiré superlattices. In these structures, the interplay between intra- and inter-layer coupling of guided modes depends strongly on the separation distance between layers\,\cite{nguyen2022magic}. Indeed, interlayer separation directly affects coupling strength, and photonic flatbands have been observed at specific "magic distances," critical separations at which light confinement is significantly enhanced and which can be effectively described by a tight-binding framework\,\cite{nguyen2022magic}. Subsequent numerical studies have explored various facets of these systems, including finite-size effects on light localization\,\cite{saadi2024many}, the robustness of flatbands\,\cite{hong2022flatband,Xia2024}, and the implementation of staggered potentials\,\cite{Trushin2025}. However, from an experimental standpoint, tuning the interlayer separation to achieve these "magic distances" is impractical, as each separation requires a distinct multilayer stack fabricated via standard planar processes. To overcome this limitation, a recent experimental study introduced an alternative strategy: tuning the moiré supercell period to achieve flatbands in structures with enlarged supercells\,\cite{jing2025observation}. While successful in generating flatbands, this approach inherently reduces the Brillouin zone size, leading to a trivial flattening of all minibands. Furthermore, as the moiré period depends on discrete numbers of period repetitions within each layer, this approach does not allow continuous observation of band flattening evolution. Consequently, an experimental demonstration of flatband tuning arising specifically from competition between intra- and inter-layer coupling in bilayer photonic crystals remains elusive.

\begin{figure*}[ht!]	\centering
\includegraphics[width=0.7\linewidth]{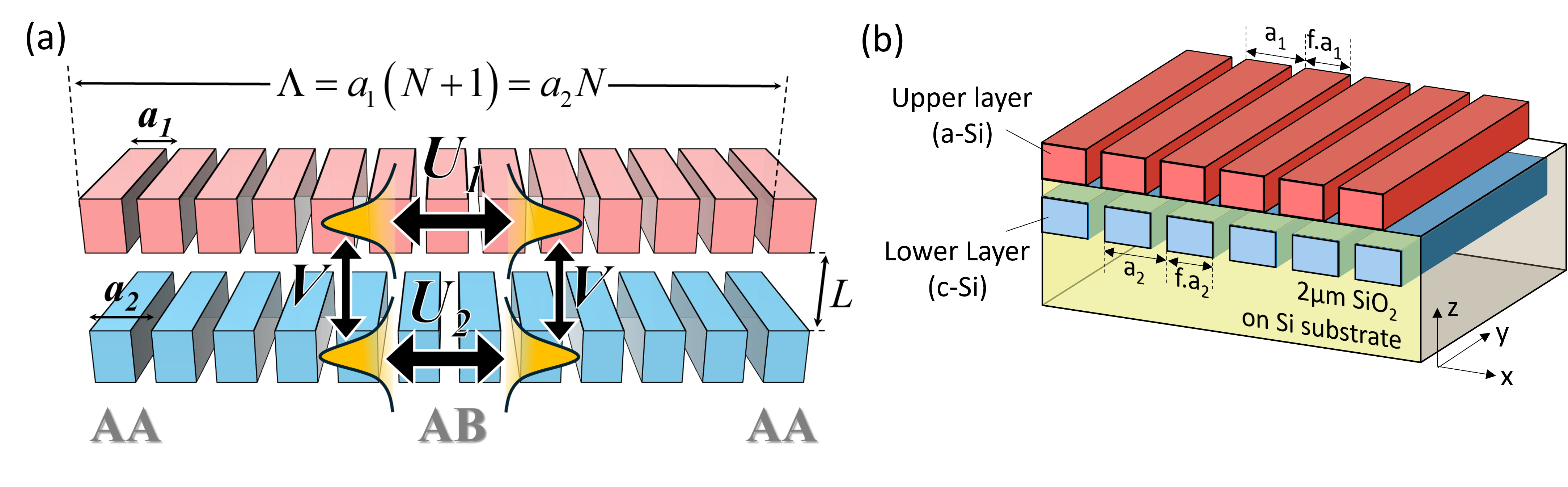} 
\caption {(a) Schematic representation of a one-dimensional (1D) moiré structure formed by two gratings with different periods $a_1$ and $a_2$, where the optical modes are governed by intralayer coupling $U_{1,2}$ and interlayer coupling $V$. The two layers are separated by a distance $L$. The moiré period is given by $\Lambda = a_1(N+1) = a_2 N$.  (b) Three-dimensional view of the structure composed of an upper layer of amorphous silicon (a-Si) and a lower layer of crystalline silicon (c-Si), deposited on a 2$\mu$m SiO$_2$ layer atop a silicon substrate. The upper and lower gratings have periods $a_1$ and $a_2$, respectively, and share the same height $h$ and filling factor $f$. The two layers are separated by a SiO$_2$ layer of thickness $L$. }
\label{fig1} 
\end{figure*}

Here, we report on the experimental dispersion engineering of moiré bands in bilayer moiré structures based on 1D silicon photonic crystals, demonstrating continuous tailoring of photonic flatbands. These bilayer structures were fabricated through a double lithography technique with high-precision alignment. The optical spacer separating the two photonic crystal layers is precisely controlled and significantly smaller than the operating wavelength, thus ensuring strong interlayer coupling. By systematically varying the refractive index contrast of the photonic crystals through adjustments in their filling factors, we effectively tuned intralayer coupling while keeping the interlayer coupling fixed by maintaining a constant interlayer spacing. Angle-resolved reflectivity measurements clearly reveal the evolution of moiré minibands in the optical dispersion diagram, arising from the competition between interlayer and intralayer coupling mechanisms. Numerical simulations using Rigorous Coupled-Wave Analysis (RCWA) strongly support our experimental observations, showing excellent agreement. Importantly, we experimentally demonstrate a flatband dispersion characterized by a significant reduction in bandwidth at a specific filling factor, consistent with theoretical predictions. These findings represent a critical step toward the precise control of photonic flatbands in bilayer moiré superlattices, offering promising applications in integrated photonics and advanced optical signal processing.

\begin{figure*}[ht!]
    \centering
    \includegraphics[width=0.6\linewidth]{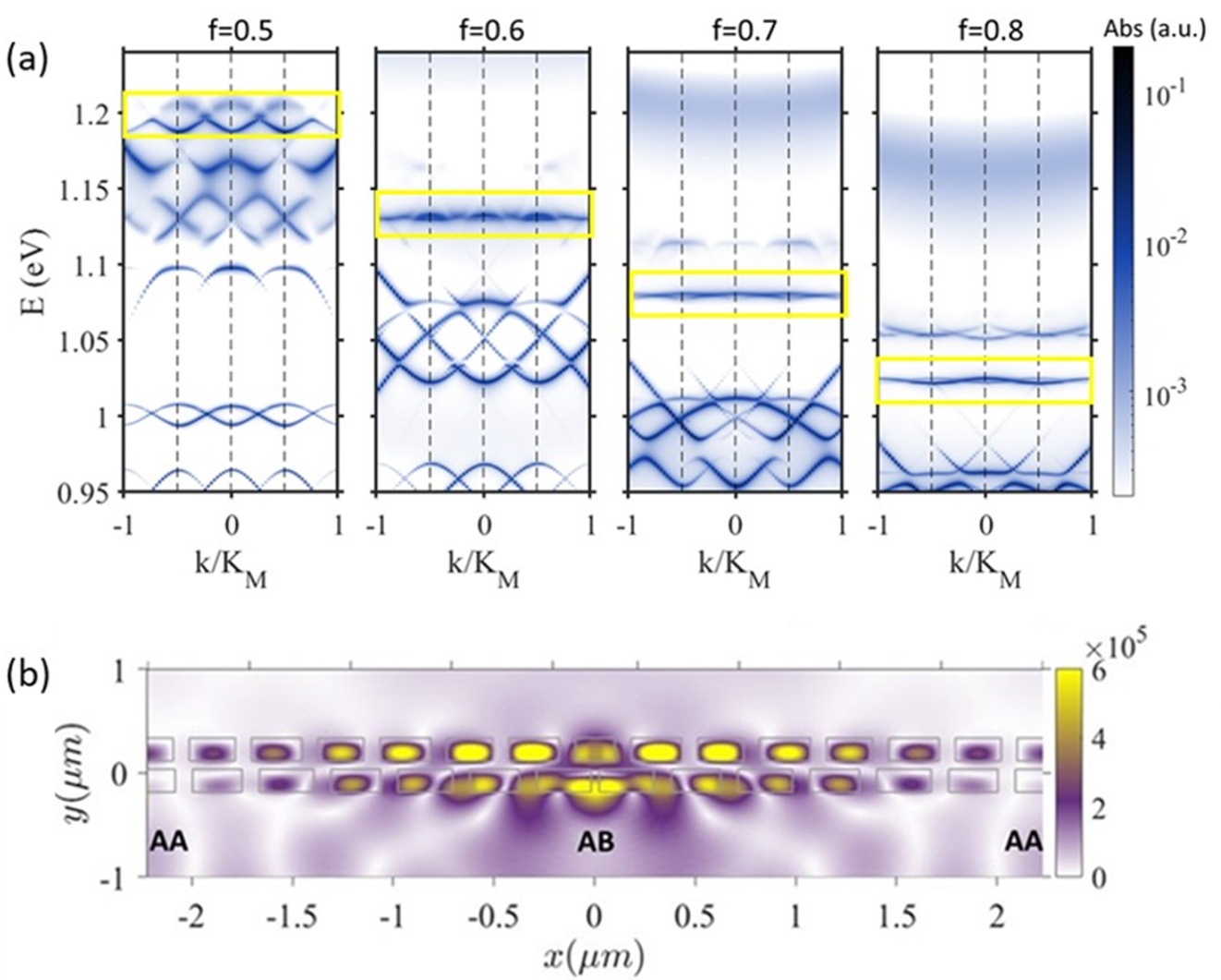}
    \caption{(a) Simulations for the moiré structure with the upper layer completely etched to 220\,nm for filling factors \( f = 0.5, 0.6, 0.7, \) and \( 0.8 \).The yellow box indicates the band of interest that varies with the filling fraction. (b) The spatial distribution of the magnetic field component $H_y$ in the moiré structure with the upper layer fully etched is shown for a filling factor $f = 0.7$ at the wavevector $k = 0$ ($\Gamma$-point), with AA and AB sites labeled.}
    \label{fig2}
\end{figure*}
\section{Emergence of Flatbands in Moiré Structures}
The studied system is a bilayer moiré photonic crystal , created by stacking two silicon photonic gratings of thickness $h$ and separated by an interlayer distance $L$ (Fig.~\ref{fig1}.a). The two gratings present slightly different lattice periods, $a_1$ and $a_2$, satisfying the commensurate conditions $\frac{a_1}{a_2}=\frac{N}{N+1}$ : $N$ is an integer which defines the size of the moiré supercell (or moiré period) $\Lambda= Na_2 = (N+1)a_1$. For each grating, the periodic pattern is defined by the same filling factor $f$, representing the proportion of high-index material in the gratings. This structure generates periodic regions at the edges of the supercells where the unit cells of the top and bottom layers are perfectly aligned (AA stacking) and regions at the centres of the supercells  where the unit cells of the upper layer are displaced relative to those in the bottom layer (AB stacking) (see Fig.~\ref{fig1}.a). These variations in stacking alter the local electromagnetic field environment and give rise to the unique characteristics of the moiré potential.

In our prior theoretical work, we showed that for a given moiré period (defined by $N$), light propagation and confinement in the structure are strongly influenced by the interplay between intra-layer and inter-layer couplings  $U_{1,2}$ and $V$ in Fig.~\ref{fig1}.a, respectively)\,\cite{nguyen2022magic}. These couplings can be tuned through structural design. The inter-layer coupling $V$, mainly governed by the overlap of the evanescent tails of co-propagating guided modes in each layer, decreases exponentially with the layer separation. In contrast, the intra-layer couplings $U_{1,2}$, arising from diffraction between counter-propagating modes within the same layer, strongly depend  on the periodic pattern and refractive index contrast of each grating. Numerical simulations have demonstrated that, in the regime of evanescent coupling, when $L<a_0$ (with $a_0=\frac{a_1+a_2}{2}$), mode hybridization between the two layers leads to the emergence of minibands within the bandgap of the uncoupled layers\,\cite{nguyen2022magic}. These minibands, analogous to those found in semiconductor superlattices, correspond to Bloch modes defined by the supercell periodicity and are subsequently referred to as moiré bands. Furthermore, adjusting the ratio between inter-layer and intra-layer coupling strengths allows the moiré band’s bandwidth to be tuned and significantly reduced. Notably, in the so-called "magic" configurations, the band becomes strongly flattened. This behavior can be effectively captured using a simple tight-binding model of an atomic chain in the Wannier basis of moiré supercells\,\cite{nguyen2022magic,saadi2024many}. Within this model, light confined in the Wannier states tunnels to nearest neighbor ones to form moiré bands.
 Notably, in the magic configurations, photon tunneling between moiré Wannier states within each supercell is canceled due to an accidental destructive interference between inter-layer and intra-layer couplings. As a result, a flatband emerges, leading to unconventional localization of photonic states within each supercell.

  Dispersion engineering of moiré bands can be achieved either by tuning the inter-layer coupling through variation of the separation distance $L$, or by tuning the intra-layer coupling via adjustment of the refractive index contrast through the filling factor $f$. In this study, we adopt the latter strategy for several reasons: firstly, varying $L$ is practically challenging, as it requires fabricating separate layer stacks for each distance, and because the inter-layer coupling decreases exponentially with $L$\,\cite{nguyen2022magic}, making fine-tuning difficult. Conversely, adjusting the filling factor $f$ can be performed continuously across different structures on the same stack, and enables precise control of the coupling strength, thanks to the linear variation of intra-layer coupling with $f$\,\cite{nguyen2022magic}.
  
  To numerically investigate the evolution of the moiré band as the filling factor $f$ varies, we calculate the energy-momentum dispersion of the modes using the RCWA method. The structure is designed to operate in the transverse magnetic (TM) mode regime around 1.2 $\mu m$ for $N=13$. To ensure a realistic design, the bilayer rests on a 2 $\mu$m silica layer  on a silicon substrate (Fig.~\ref{fig1}.b). The lower grating is composed of crystalline silicon (c-Si) and silica with a period of $a_2=317$ nm, while the upper grating consists of amorphous silicon (a-Si) and air with a period of $a_1=342$ nm.  A double-period perturbation of 10\,$\%$, similar to the designs in\,\cite{nguyen2018symmetry,nguyen2022magic}, is implemented for each grating so that the modes of interest, originally near the X point of the BZ, are brought to the center of the BZ and can be probed by plane-wave excitations in RCWA.
The thickness  $h$ of each grating is $220$~nm. The two gratings are separated by a $SiO_2$ layer fixed at $L=80$ nm to ensure the evanescent coupling condition ($L<a_0$ with $a_0=\frac{a_1+a_2}{2}\approx 330$~nm). The refractive indices of all materials used in the simulations were obtained from ellipsometry measurements. In the spectral range of interest, they are: \( n_{\text{a-Si}} = 3.53 \), \( n_{\text{c-Si}} = 3.51 \), \( n_{\text{SiO}_2\,(\text{SOI})} = 1.45 \), and \( n_{\text{SiO}_2\,(\text{spacer})} = 1.44 \). The filling factor $f$ varies from 0.5 to 0.8 to enable fine-tuning of the intra-layer couplings.

 Figure~\ref{fig2} presents numerical results highlighting the evolution of photonic bands as the filling factor $f$ varies from 0.5 to 0.8, with emphasis on the isolated moiré band (marked in yellow). At lower filling factors (\( f = 0.5 \) and \( f = 0.6 \)), clear and distinct moiré bands emerge within the band structure, indicating new energy states introduced by the moiré corrugation (Fig.~\ref{fig2}.a). As the filling factor increases to \( f = 0.7 \), the moiré band exhibits significant flattening, reflecting a balance between interlayer and intralayer coupling. However, further increasing the filling factor to \( f = 0.8 \) causes the band to become dispersive again, suggesting that the minimum bandwidth (flatband condition) is achieved at a filling factor close to 0.7.The band flattening has a direct impact on the spatial distribution of the electromagnetic field in the structure.  Figure.~\ref{fig2}.c shows the computed map of the field intensity ($|H_y|^2$) for a filling factor of $f=0.7$, corresponding to the flatband configuration. The results clearly illustrate strong field localization at AB sites, whereas AA sites exhibit comparatively weaker confinement. These findings confirm that the moiré flatbands arise from the intricate interplay of interlayer and intralayer interactions and are tunable via the filling factor \( f \).

\begin{figure*}[hbt!]
    \centering
    \includegraphics[width=0.7\linewidth]{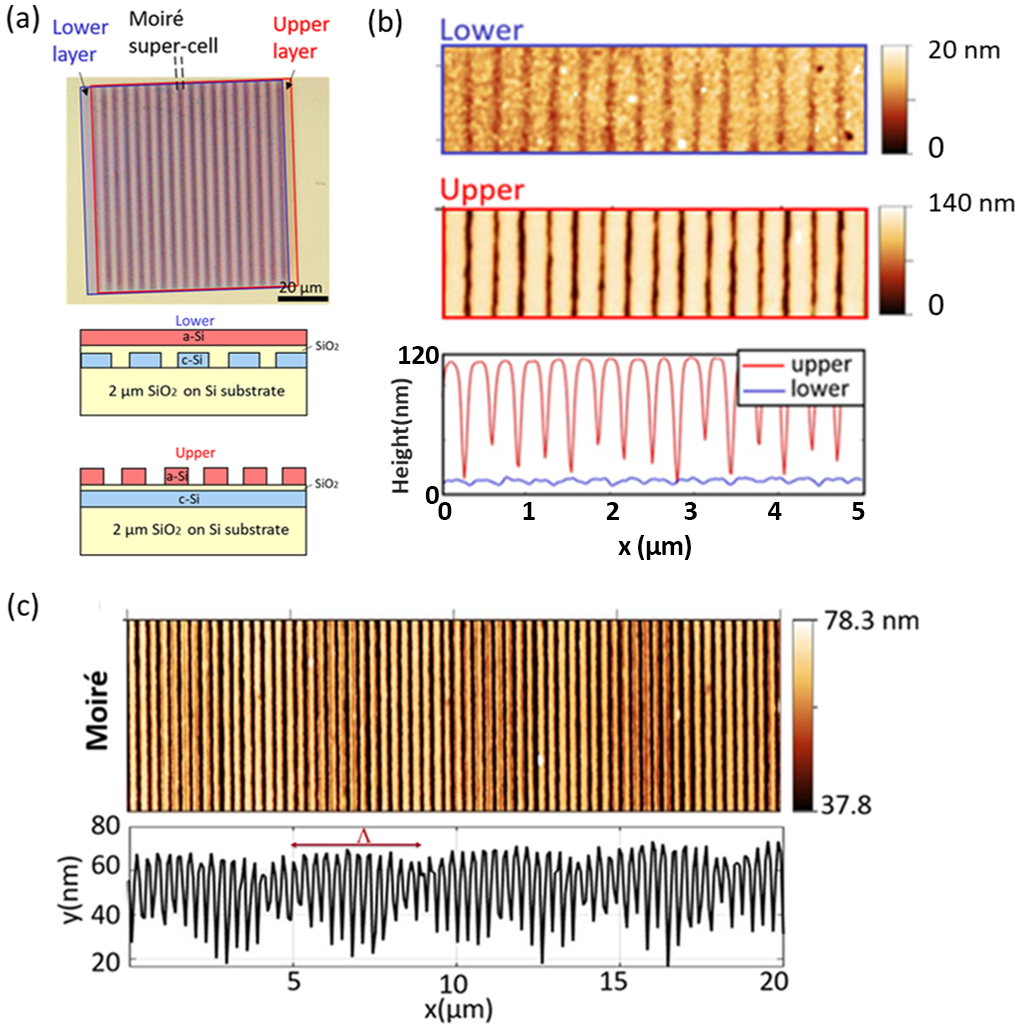}
    \caption{(a) Image captured by an optical microscope showing the formation of a moiré supercell from the superposition of two layers. The lower layer (outlined in blue) and the upper layer (outlined in red) create an interference pattern visible as the moiré supercell. (b) Height profiles of the upper and lower layers are shown separately. Below the profiles, a combined plot illustrates the periodic variations in height across a 5 µm span, with the red and blue lines representing the upper and lower layers, respectively. (c)  Height profile of the moiré supercell structure is shown. The profile illustrates the periodic modulation of the moiré pattern over a \(20 \, \mu \text{m}\) span. The lower plot provides the height variation along the x-axis, highlighting the periodic nature of the moiré interference pattern, with the moiré periodicity denoted as \(\Lambda \sim 4 \, \mu \text{m}\).}
    \label{fig3}
\end{figure*} 

\section{Fabrication and Characterization of Moiré Photonic Crystals}

To fabricate the moiré sample, we employed a monolithic  fabrication process involving double nanostructuring with precise alignment, enabling accurate definition of photonic crystal layers. These structures were patterned using conventional planar microelectronics techniques. The process began by nanopatterning the lower gratings from a commercial 220 nm silicon-on-insulator (SOI) substrate (Fig.~\ref{fig3}.a).Multiple gratings, each with the same period $a_2=342$ nm but different filling factors $f$ were defined. The size of each grating is $80$ µm × $80$ µm. These patterns were initially created in an $80$ nm thick AR-P resist layer using electron beam lithography, then transferred to the c-Si layer via ionic dry etching. Subsequently, a two-phase planarization process was performed: first, a silica (SiO\(_2\)) layer was deposited by a sol-gel process, followed by controlled ionic dry etching precisely planarize the surface and achieve the desired $80$ nm spacer between the two gratings. Ensuring this uniform surface is crucial for accurate control over the optical properties and to minimize irregularities that could affect device performance.

The upper grating, consisting of a-Si gratings (\( h = 220 \) nm) deposited by plasma-enhanced chemical vapor deposition (PECVD), was fabricated through a second electron beam lithography step, accurately aligned with the lower grating to form the moiré pattern. Precise alignment  is essential, as a tilt between the two gratings could cause significant misalignment, impacting device efficiency. Regarding the lateral alignment, numerical simulations confirmed that horizontal shift up to 20 nm  between the grating rods has negigible effects on  optical frequencies when the number of moiré periods $N$ is sufficiently large ($N \gg 1$) (see supporting information). Moreover, even larger global shift of a few micrometers  does not affect the optical performance due to the inherent periodicity of the moiré pattern.

All lower grating structures have a period of \(a_2 = 342 \) nm with filling factors ranging from \( f = 0.53 \) to \( 0.78 \), while the upper grating has a period of \(a_1 = 317 \) nm with similar filling factors (for more details on the fabrication, see the Supporting Information). A double-period perturbation of 10$\%$ is also implemented in the design of the fabricated structures, allowing the photonic modes to be probed via far-field spectroscopy (angle-resolved reflectivity).
  \begin{figure*}[ht!]
	\centering
\includegraphics[width=0.7\linewidth]{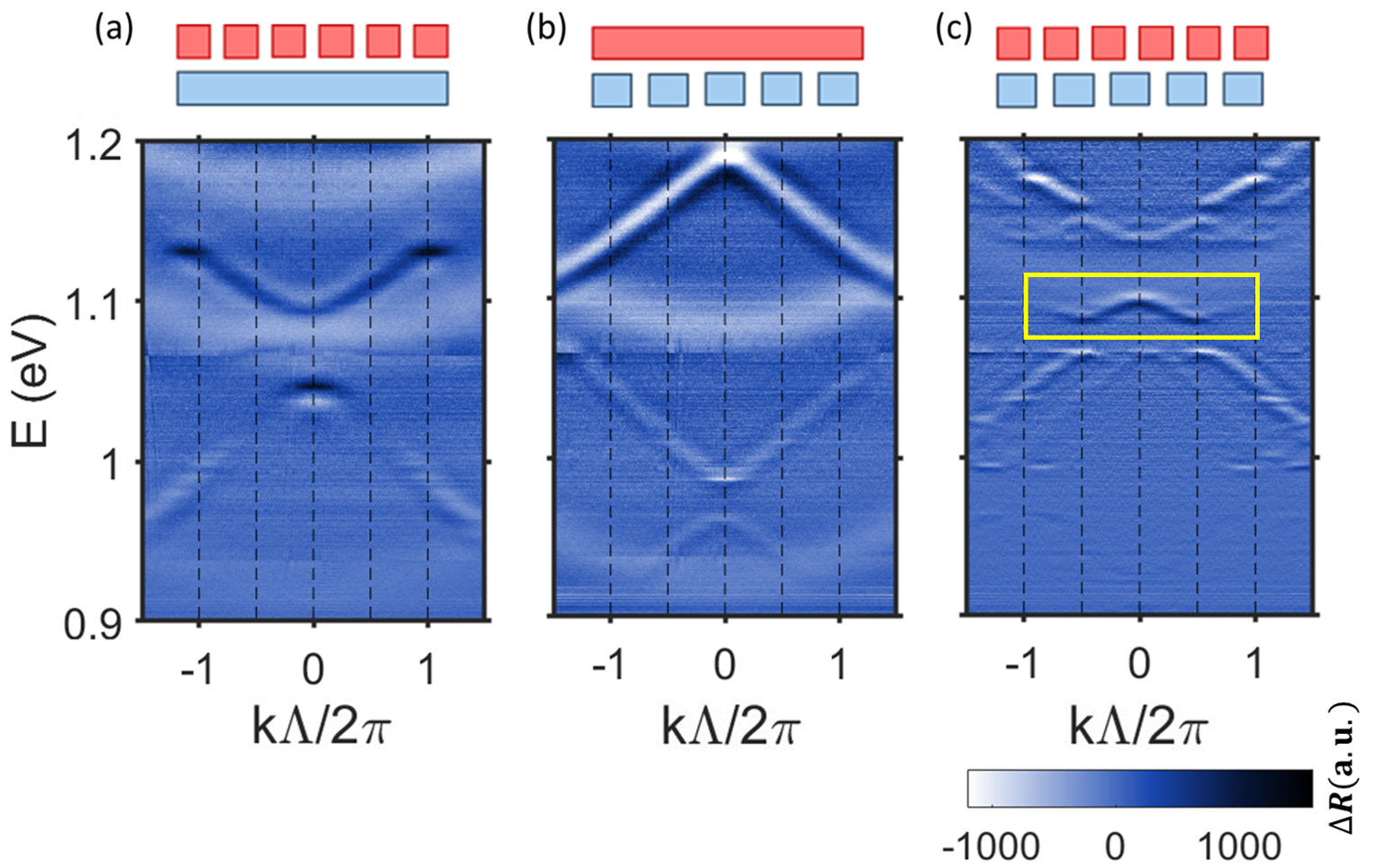} 
\caption{Experimental dispersion diagrams obtained from angle-resolved reflectivity measurements at different regions of the bilayer moiré photonic crystal. The maps show the filtered differential reflectivity signal \( \Delta R \) (a.u.); see Supporting Information for details. (a) Dispersion of the upper grating, (b) dispersion of the lower grating, and (c) dispersion in the moiré region, where moiré bands emerge due to interlayer interactions. The yellow box highlights the isolated moiré flatband observed around \( E = 1.1\,\text{eV} \).}
\label{fig4}
\end{figure*}

A microscopic image was captured to access the fabrication quality of the moiré structure (Fig.\ref{fig3}.a, top panel). This image clearly shows the superposition of the two membranes with negligible tilt: the lower layer, outlined in blue, and the upper layer, outlined in red, with the resulting moiré supercell measuring approximately 4 $\mu m$, closely matching the theoretical prediction given by $\Lambda = 4.4$ $\mu m$. It is important to point out that due to the positive nature of the resist,  outside of the overlap region, both lower and upper stack is still a bilayer stack, but only one layer is patterned :  lower stack consists of an unpatterned a-Si slab on top a c-Si grating, and upper stack consists of an a-Si grating on top of an unpatterned c-Si slab (Fig.\ref{fig3}.a, lower panels).  

To assess the quality of planarization atomic force microscopy (AFM) measurements of the lower stack, taken over a span of $5$ $\mu m$ were performed. The result (Fig.\ref{fig3}.b), top panel, show a modulation depth of approximately $6$ nm, confirming successful planarization from an initial modulation depth of $220$ nm of the patterned c-Si grating in air down to $6$ nm. Moreover, ellipsometry measurements outside the patterned regions (see supplemental material) further confirmed the precise thickness of the silica spacer, validating the accuracy of the etching process in the planarization step.  AFM analysis of the upper stack indicates a modulation depth of around $130$ nm ,smaller than the total a-Si thickness of 220 nm (Fig.\ref{fig3}.b). This partial etching likely resulted from altered etching dynamics due to the presence of the lower grating. Nonetheless, as demonstrated later, this partial etching minimally affects dispersion engineering and flatband formation in the moiré structures. Finally, AFM analysis of the moiré structure over a $20$ $\mu m$ span reveals that a residual modulation caused by the planarization step results in a topographic beat of approximately $6$ nm (Fig.\ref{fig3}.c). Such beating, of period 4.4 $\mu m$, consistent with the moiré period identified in the microscopic image (Fig.\ref{fig3}.a), and alinging well with theoretical prediction.

\section{Experimental Measurement  of Moiré Flatbands}
To measure the band diagram of our structure, a white light beam is focused onto the sample via a microscope objective (NA=$0.8$). The reflectivity is collected using the same microscope in a confocal geometry and is then projected into Fourier space before being directed to the entrance slit of a spectrometer. Finally, the signal is captured by the sensor of an infrared camera at the spectrometer output\,\cite{cueff2024fourier} (see Supporting Information).  

Optical measurements were performed in three distinct regions to analyze the response of the three different stacking configurations: the upper stacking, the lower stacking, and the moiré region (Fig.~\ref{fig4}).
The dispersion diagrams corresponding to the optical response of the upper and lower stackings exhibit the well-known characteristic 1D photonic crystal band structures (Fig.~\ref{fig4}.a-b), albeit slightly modified with respect to the band structures of single gratings due to the presence of the layered stack  (see Supporting Information for RCWA simulations of the upper and lower structures). In contrast, the moiré region exhibits a markedly different optical response compared to the individual gratings. Interlayer interactions between the two gratings induce significant modifications in the photonic dispersion, leading to the emergence of moiré bands. Experimental results reveal the formation of distinct moiré bands, notably an isolated band (highlighted in yellow) around E = 1.1 eV, which is the band of interest in this study (Fig.~\ref{fig4}.c).

\begin{figure*}[ht!]
	\centering
\includegraphics[width=\linewidth]{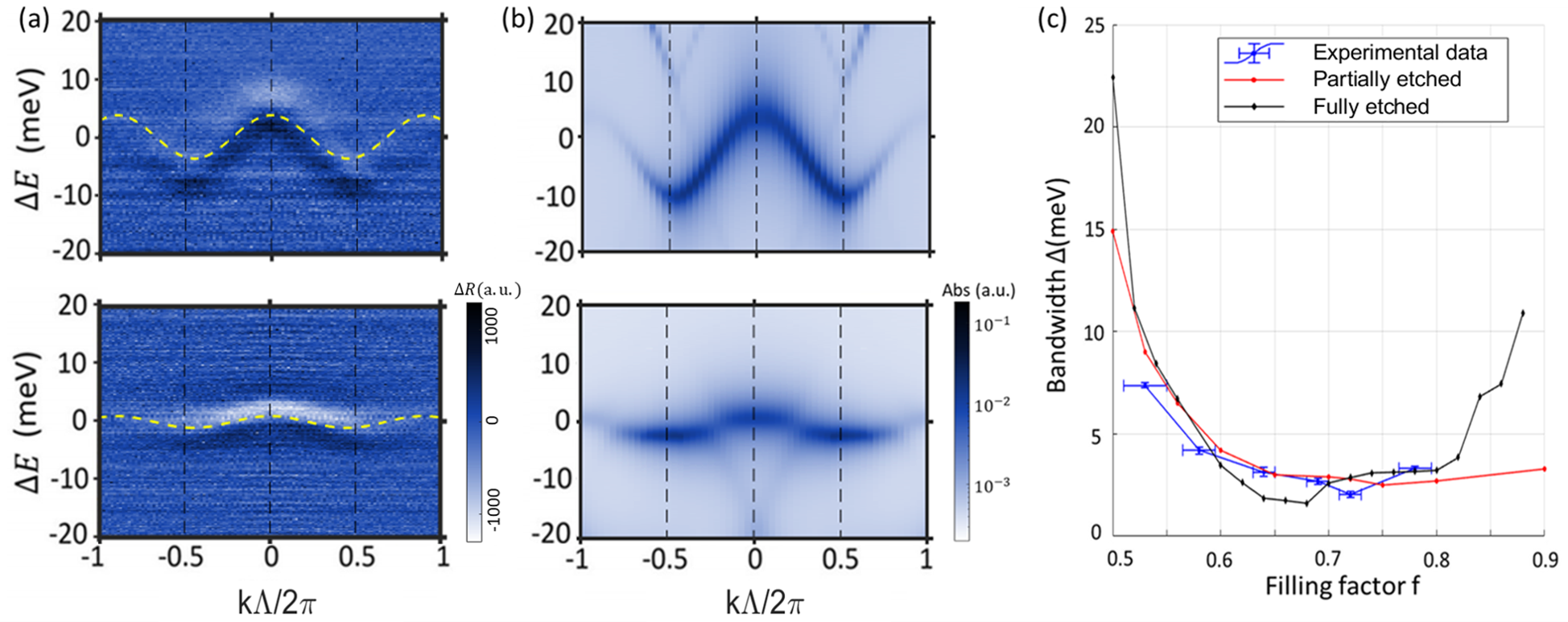} 
		\caption{(a) Experimental measurements for \( f = 0.53 \) and \( f = 0.72 \), with cosine fits (yellow lines) derived from the model equation. 
(b) Corresponding RCWA simulations for \( f = 0.53 \) and \( f = 0.75 \), showing good agreement with the positions of the experimental photonic bands. These results highlight a progressive transition toward a flatband for \( f = 0.72 \). (c) Variation of the bandwidth \( \Delta \) (meV) as a function of the filling fraction \( f \). The experimental measurements in blue were performed for \( f = 0.53 \), \( 0.58 \), \(0.64 \), \( 0.69 \), \( 0.72 \), and \(  0.78 \). The data points (in blue) are displayed with error bars. The RCWA simulations are shown in red and black.}
\label{fig5}
\end{figure*}

This band highlighted in yellow evolves as the filling factor $f$ is progressively tuned, as shown in  Fig.~\ref{fig5}. At $f = 0.53$, the band exhibits noticeable dispersion. As $f$ increases, a transition toward minimal dispersion occurs, leading to the formation of a flatband around $f = 0.7$. At $f = 0.72$, the experimental data clearly reveal a significantly flattened band, confirming the emergence of a moiré flatband.

To further support this observation, we compare our experimental findings with RCWA simulations 
of the fabricated structures with partially etched upper grating (Fig.~\ref{fig5}.b). The simulated energy-momentum dispersion closely matches the experimental band positions and their evolution as \(f\) increases, validating the robustness of our fabrication and measurement approach.  
Theoretical simulations predict that the formation of flatbands in these structures is closely tied to variations in the filling factor \(f\). In a previous study\,\cite{nguyen2022magic}, this behavior was modeled using a tight-binding approach that considers the nearest-neighbor couplings. In this model, the moiré structure is represented as a periodic chain of supercells, where the energy profile between \(k = 0\) and \(k = \pi/\Lambda\) (corresponding to the \(\Gamma\) and \(X\) points in the first Brillouin zone) is given by:  

\begin{equation}
    E(k) = E_0 + 2J\cos(\Lambda k),
    \label{EQ}
\end{equation}

where \(E_0\) is the central energy of each supercell, \(J\) is the tunneling rate governing the coupling between neighboring supercells, and \(\Lambda\) is the moiré period. The global spectral bandwidth \(\Delta\), defined as \(\Delta = 4J\), serves as a figure of merit to evaluate the flatness of the moiré minibands. A smaller bandwidth \(\Delta\) corresponds to a flatter band, indicating minimal dispersion.  In this equation \label{E}, the energy deviation relative to the central energy \(E_0\) is given by \(\Delta E = E(k) - E_0\), where \(\Delta E\) characterizes the energy variation within the band.  
In Fig.~\ref{fig5}.a, the yellow lines represent cosine fits derived from the tight-binding model (Eq.~\ref{EQ}). These fits closely match the experimental band dispersion, confirming the periodic modulation of energy.

Additionally, we present the evolution of the bandwidth \(\Delta\) as a function of the filling factor \(f\) (Fig.~\ref{fig5}.c) to further quantify this behavior. The graph compares three datasets: experimental measurements (blue curve), RCWA simulations for a partially etched upper layer (red curve), and RCWA simulations for a fully etched upper layer (black curve). The experimental results (blue) show a clear trend where the bandwidth \(\Delta\) decreases with increasing \(f\), reaching a minimum of approximately \(2 \, \text{meV}\) at \(f = 0.72\), where the band is nearly flat. For \(f > 0.72\), the bandwidth increases again, indicating a gradual return to a more dispersive state.
The RCWA simulations of both partially etched and fully etched configurations closely follow the experimental data, showing a sharp decrease in \(\Delta\) from \(f = 0.5\) to a minimum near \(f = 0.7\). These findings demonstrate that the filling factor \(f\) is a critical tuning parameter for controlling the flatness of moiré bands. Despite small discrepancies between the experimental and simulated values—likely due to fabrication-induced variations in the filling factor and etching depth—the general agreement highlights the robustness of the observed behavior. The emergence of a flatband near \(f = 0.72\) confirms the theoretical simulations and emphasizes the potential of moiré structures for achieving tunable photonic states with minimal dispersion.  The measured bandwidth remained approximately 2 meV, indicating that while the band is relatively narrow, it is not entirely flat. The ratio of the bandwidth to the central energy ($\Delta/E_0$) was found to be approximately 1:500, highlighting the relatively small bandwidth in comparison to the energy scale.  

\section{Conclusion and Perspectives}
This study presents the experimental observation of flatbands in one-dimensional moiré photonic crystal structures. By combining RCWA  simulations with precise experimental techniques, we confirmed the emergence of flatbands within these structures. The results align well with theoretical expectations, particularly in demonstrating minimal dispersion at specific filling factors. By adjusting key parameters, such as the filling factor, we achieved fine-tuning of intralayer coupling, leading to enhanced light confinement and the formation of localized photonic states. The fabrication process that we developed can also enable the realization of finite-size cavities, opening a promising route toward  compact, high-Q resonators\,\cite{saadi2024many}. Furthermore, this versatile platform can be extended to investigate topological phenomena in multilayer photonic crystals\,\cite{Lee2022,Nguyen2023,NguyenThoulessPumpping}, offering new opportunities for advanced studies of light–matter interactions.

Looking forward, our platform offers exciting prospects beyond passive photonic structures. By incorporating gain media such as III-V semiconductors or 2D materials between the gratings, moiré-based laser cavities could be realized. Furthermore, the strong sensitivity of the flatband regime to the filling factor makes these structures excellent candidates for optical sensing applications, with the added advantage of broad angular acceptance for detecting gases or liquids.

\section*{acknowledgements}
This work was supported by Nanolyon platform, a member of the CNRS-RENATECH+ French national network. The authors thank Pierre Viktorovitch, Serge Mazauric and Xuan Dung Nguyen for fruitful discussions. We acknowledge support from the CNRS/IN2P3 Computing Center (Lyon - France) for providing computing and data-processing resources needed for this work.

\bibliography{main.bib}

\setcounter{equation}{0}
\setcounter{figure}{0}
\setcounter{table}{0}

\setcounter{equation}{0}
\setcounter{figure}{0}
\setcounter{table}{0}
\setcounter{section}{0}

\renewcommand{\theequation}{S\arabic{equation}}
\renewcommand{\thefigure}{S\arabic{figure}}
\renewcommand{\bibnumfmt}[1]{[S#1]}
\renewcommand{\vec}[1]{\boldsymbol{#1}}

\onecolumngrid 
 \begin{center}
\Large{--- SUPPLEMENTAL INFORMATION ---}
\end{center}
\normalsize{}
\section{Fabrication of the 1D moiré}
\label{Fabrication of the 1D moiré}
For the fabrication, we used the technique of monolithic stacking of successive layers to fabricate 1D moiré through double nanostructuring with alignment. This approach is based on classic planar microelectronics techniques such as lithography and etching. The patterns of the photonic crystals must be superimposed in a controlled manner to obtain a 1D moiré structure. The photonic structures must be superimposed with a lateral alignment that must be a maximum of $80$ $\mu m$ and a distance between the membranes that must be less than $100$ nm to observe the expected effects. We present the development of this technological process to create periodic or aperiodic patterns with slightly different periods, leading to moiré effects.
The fabrication of a structure includes several important steps:
\begin{enumerate}
\item Fabrication of the lower photonic crystal using electron beam lithography (e-beam) and ICP-RIE etching  (fig.\ref{S1}).
\begin{figure}[ht!]
    \centering
\includegraphics[width=\linewidth]{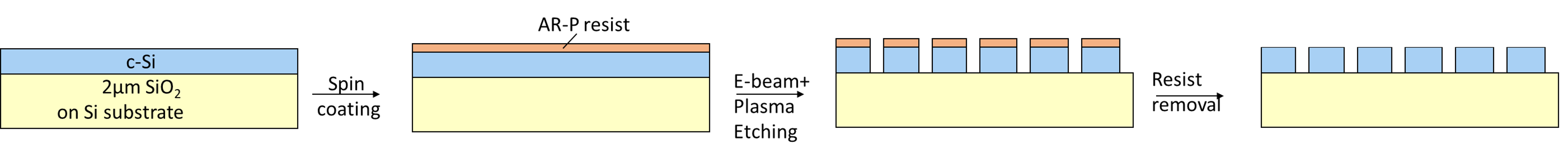}
    \caption{Fabrication of the lower photonic crystal.}
    \label{S1}
\end{figure}
\item Planarization of the surface and deposition of silica to control the distance between the two photonic crystals  (fig.\ref{S2}).
\begin{figure}[ht!]
    \centering
\includegraphics[width=0.9\linewidth]{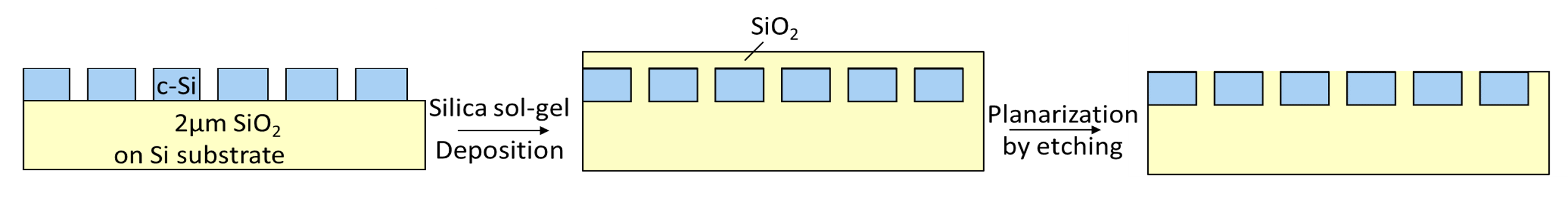}
    \caption{Planarization of the surface and deposition of silica to control the distance between the two photonic crystals. The process involves silica sol-gel deposition and planarization by etching to achieve the desired thickness.}
    \label{S2}
\end{figure}
\item Deposition by PECVD (plasma enhanced chemical vapor deposition) of the amorphous silicon layer for the realization of the second photonic crystal.
\item Fabrication of the upper photonic crystal with precise alignment relative to the lower photonic crystal to form the moiré pattern  (fig.\ref{S3}).
\end{enumerate}
\begin{figure}[ht!]
    \centering
\includegraphics[width=\linewidth]{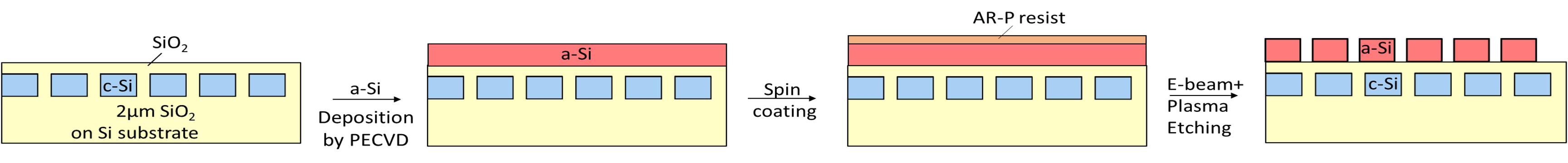}
    \caption{Fabrication of the upper photonic crystal with precise alignment relative to the lower photonic crystal to form the moiré pattern.}
    \label{S3}
\end{figure}
The development efforts mainly focused on steps (2) and (4). The main challenge lies in mastering the silica deposition and planarization process to control the distance between the two photonic crystals. Additionally, it is crucial to adjust the relative vertical and horizontal positions of the photonic crystals (PC) to ensure alignment during the fabrication of the upper photonic crystal. Besides, it is particularly crucial to minimize angular tilt during alignment.Note that the photonic modes of uncoupled gratings are positioned below the light-line, and without sufficient inter-layer coupling, these modes would not be accessible for angle-resolved reflectivity measuremnts. To address this, we introduce a 5\% double period perturbation in the design of each grating for the fabrication. The design uses unit cells consisting of two moiré patterns: one compressed to \( 0.95\Lambda \) and the other expanded to \( 1.05\Lambda \).

\section{Ellipsometric Characterization of the Multilayer Stack}
\begin{figure}[ht!]
    \centering
    \includegraphics[width=0.8\textwidth]{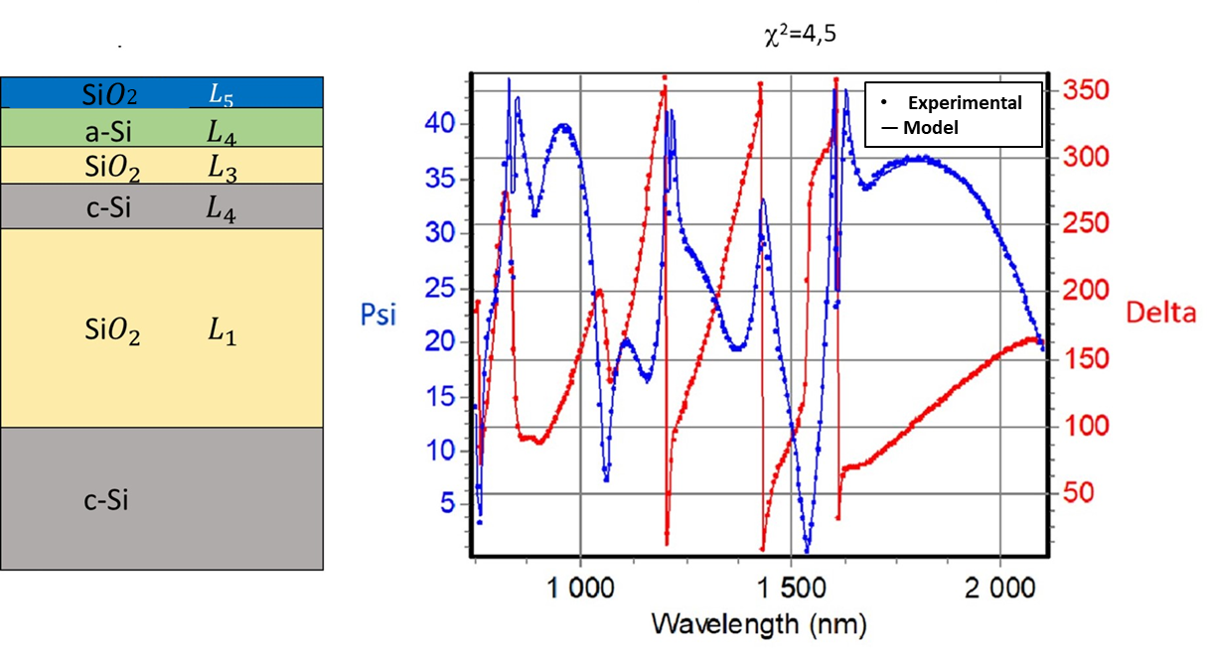}
    \caption{(Left) Schematic representation of the layer stack for sample E2, consisting of a SiO\textsubscript{2} hard mask (\(L_5\)), an a-Si layer (\(L_4\)), an intermediate SiO\textsubscript{2} layer (\(L_3\)), a c-Si layer (\(L_2\)), and a base SiO\textsubscript{2} layer (\(L_1\)) on a silicon substrate. (Right) Ellipsometric measurements of \(\Psi\) (blue) and \(\Delta\) (red) as functions of wavelength. Dots correspond to experimental data, and solid lines represent the fitted model. The fitting was performed by varying the thicknesses of \(L_3\), \(L_4\), and \(L_5\), resulting in a good agreement with experimental results (\(\chi^2 = 4.5\)).}
    \label{fig:ellipsometry}
\end{figure}

To verify that the layer thicknesses of the stack matched the nominal values, ellipsometric measurements were performed. For sample E2, the stack consists of SiO\textsubscript{2}/c-Si/SiO\textsubscript{2} (sol-gel)/a-Si/hard mask. Figure~\ref{fig:ellipsometry} illustrates the layered structure, from top to bottom: the SiO\textsubscript{2} hard mask, the a-Si layer, the intermediate SiO\textsubscript{2} layer, a c-Si layer, all deposited on a SiO\textsubscript{2} base layer on a silicon substrate. The corresponding thicknesses are labeled from \(L_1\) to \(L_5\). The measurements reveal some deviations from the nominal values. For instance, the \(L_5\) SiO\textsubscript{2} mask layer was measured at 53.3~nm instead of the expected 45~nm, while the \(L_4\) a-Si layer thickness was found to be 237~nm versus 220~nm. The \(L_3\) SiO\textsubscript{2} layer measured 79.3~nm (nominal value: 80~nm), and the \(L_2\) c-Si layer was 215~nm thick. Finally, the bottom \(L_1\) SiO\textsubscript{2} layer had a thickness of 2.002~\textmu m.

Figure~\ref{fig:ellipsometry} also shows the experimental ellipsometric parameters Psi and Delta, along with the values obtained by fitting a five-layer model. In this model, the fitting procedure involved varying the thicknesses of layers \(L_3\), \(L_4\), and \(L_5\), while keeping the other parameters fixed. This approach provided an optimal match between the experimental data and the modeled response. \(\Psi\) represents the amplitude ratio of the p- and s-polarized reflection coefficients, while \(\Delta\) corresponds to the phase difference between these two components. Despite the complexity of the multilayer system, the fit quality is good, enabling an unambiguous determination of the opto-geometrical parameters of the stack.

\newpage
    \label{supp8}
\section{Lateral displacement between photonic crystal rods}

To demonstrate that a 20 nm displacement between the rods during the fabrication of photonic crystals does not affect the optical response of the structure, we simulated asymmetric structures by varying the period $a_1$ such that $a_1 = a_1 + 20$ nm using COMSOL Multiphysics. The band diagrams corresponding to this configuration indicate that there is no significant change in the bandgap or energy levels , confirming that the optical response of the photonic crystal structure remains insensitive to this fabrication-induced displacement.

\begin{figure}[ht!]
    \centering
    \includegraphics[width=0.9\linewidth]{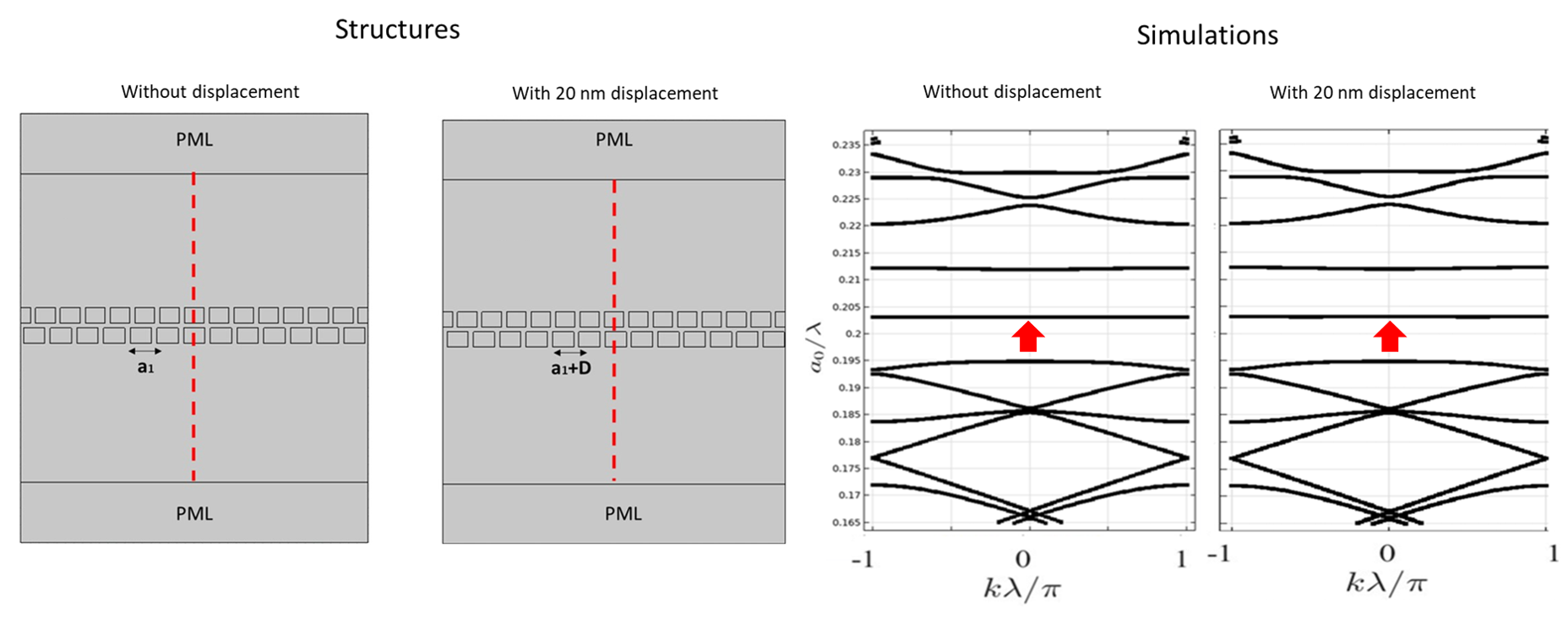}
    \caption{Comparison of the photonic crystal structures and their band diagrams for the cases without displacement (left) and with a 20 nm displacement (right). The red arrows highlight a specific flatband, showing that its position remains unchanged despite the displacement, confirming the robustness of the optical response.}
    \label{fig:setup}
\end{figure}
\newpage
\section{Robustness and Tunability of Moiré Flatbands }
To evaluate the robustness of these flatbands, we simulated the structure with an upper layer partially etched to a thickness of 130 nm. Despite this modification, the results (Fig. 2.b) confirm that moiré flatbands remain intact, though they exhibit an energy shift and appear at a slightly different filling factor. This behavior highlights the influence of etching depth on the overall dispersion properties of the system. These findings substantiate that the moiré flatbands originate from the intricate coupling between interlayer and intralayer interactions and are tunable via the filling factor f . Furthermore, their invariance under thickness variations underscores the practical feasibility of realizing moiré-based photonic structures with stable flatband characteristics.
\begin{figure}[ht!]
    \centering
    \includegraphics[width=0.92\textwidth]{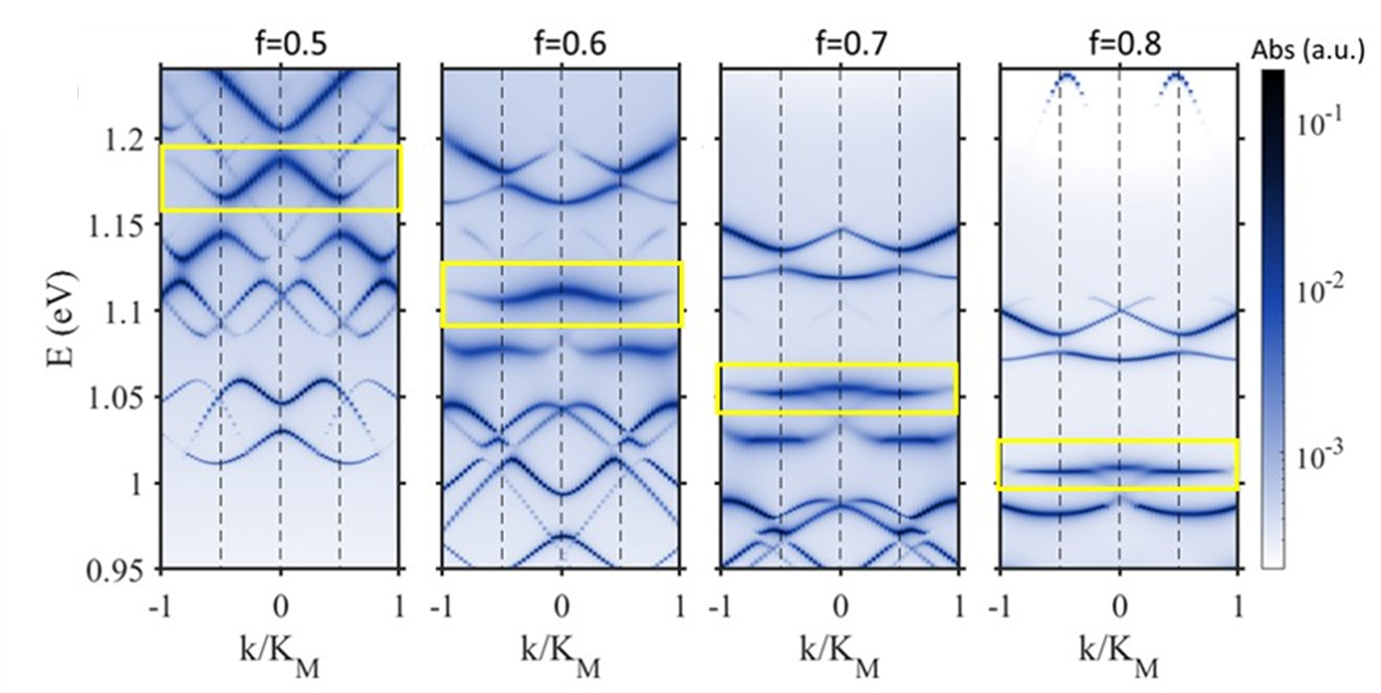}
    \caption {Simulated photonic band structures for a moiré superlattice with varying filling factors $f = 0.5$, $0.6$, $0.7$, and $0.8$ in the case of a partially etched upper layer with a thickness of 130 nm.}

    \label{fig:grating_dispersion}
\end{figure} 

\newpage
\section{Dispersion for the upper and lower layers}
We compare experimental and simulated dispersion diagrams of photonic crystal of the upper and lower layers. The results show good agreement with RCWA simulations, accurately capturing key features of the photonic band structures.

\begin{figure}[ht!]
    \centering
    \includegraphics[width=0.5\textwidth]{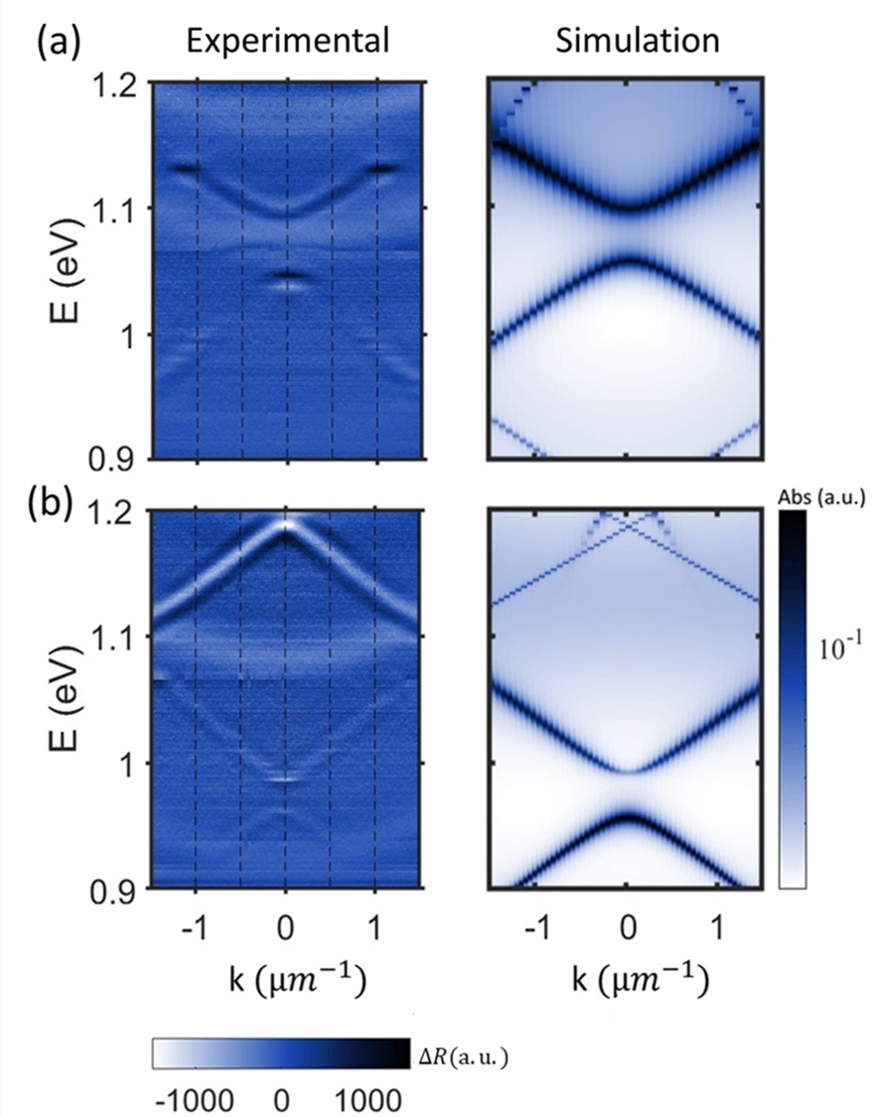}
    \caption{(a) Dispersion of the upper layer (thickness: 130 nm): experimental measurement (left) and corresponding RCWA simulation (right).  (b) Dispersion of the lower layer (thickness: 220 nm): experimental measurement (left) and corresponding RCWA simulation (right).  
    }
    \label{fig:grating_dispersion}
\end{figure}

\section{Angle-resolved reflectivity setup}
The angle-resolved reflectivity measurements presented in Fig. \ref{fig:setup} of the main text were conducted using a custom-built Fourier setup (illustrated in Fig. \ref{fig:setup}). A halogen lamp provides illumination, and a 0.8 NA microscope objective is used to both illuminate the sample and collect the reflected light. This reflected light is then directed through a series of lenses and into a spectrometer, where it is captured by a  camera. The Fourier plane of the sample, which is located at the Back Focal Plane (BFP) of the objective, is projected onto the entrance slit of the spectrometer using Fourier and focus lenses. The dispersion along the x-axis is selected by the spectrometer slit, diffracted by the spectrometer’s diffraction grating, and finally projected onto the CCD camera, resulting in a kx-$\lambda$ dispersion map.

\begin{figure}[ht!]
    \centering
    \includegraphics[width=1\linewidth]{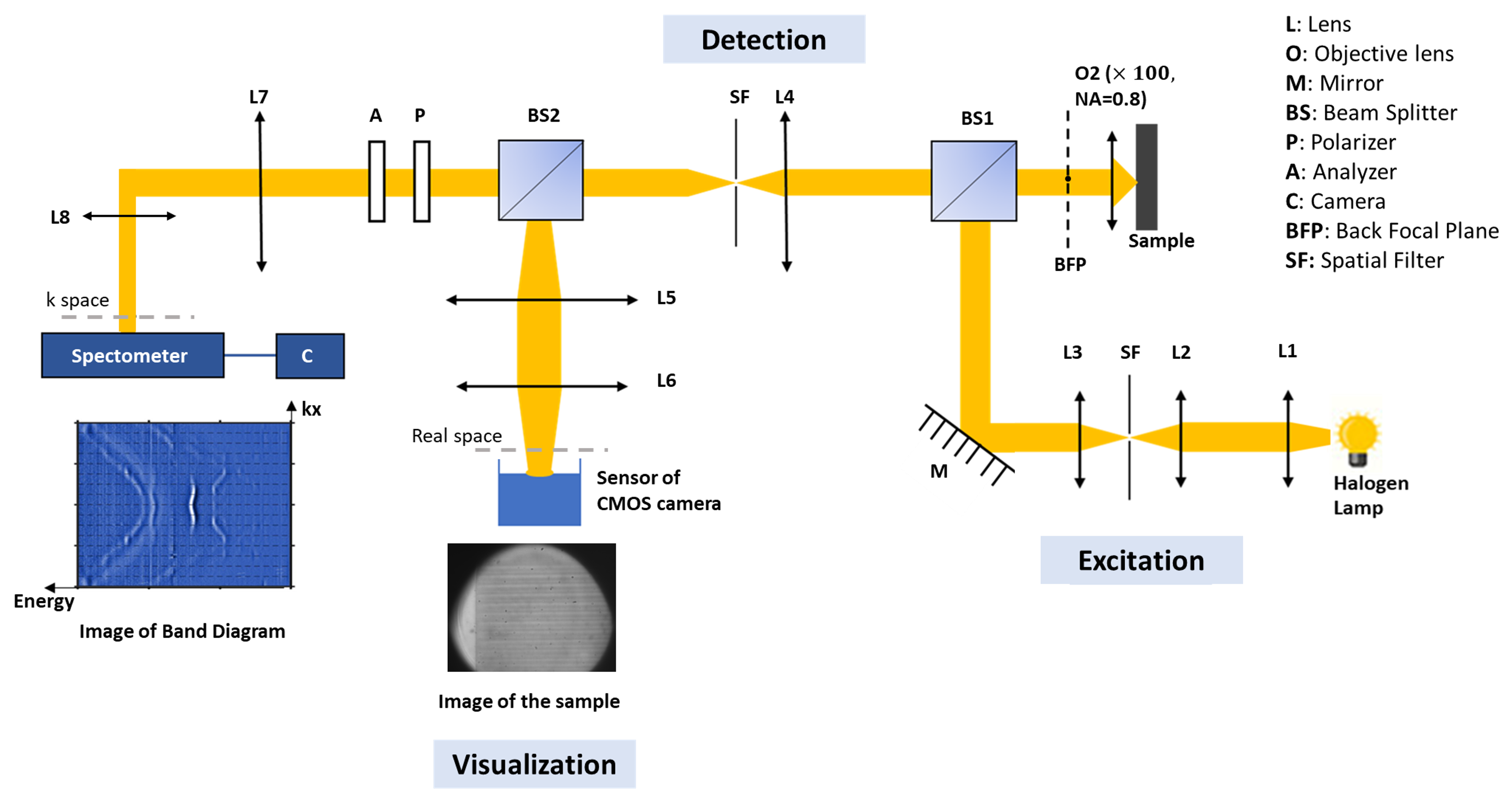}
    \caption{Schematic of the custom-built Fourier setup used for reflectivity measurements.}
    \label{fig:setup}
\end{figure}
\newpage
\section{Filtered parasite modes method}

\begin{figure}[ht!]
    \centering
\includegraphics[width=0.8\linewidth]{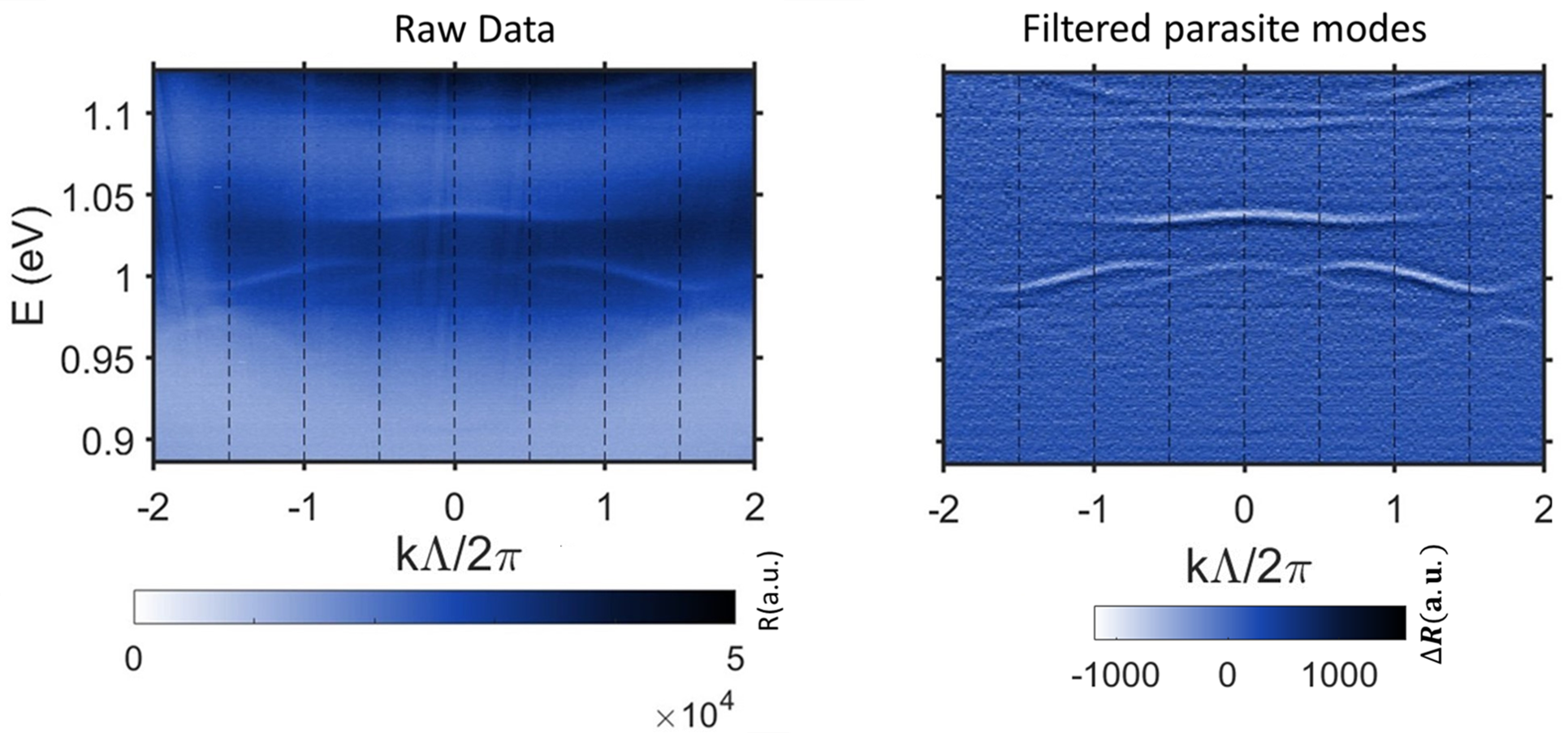}
    \caption{The left panel shows the raw data, displaying the unprocessed reflectivity spectrum, where the vertical axis represents the energy \( E \) (in eV) and the horizontal axis represents the wave vector \( k \) (in \( \mu \text{m}^{-1} \)). 
The right panel shows the filtered data, obtained after applying the filtering method, which reveal the characteristics of the photonic mode.
}
    \label{S5}
\end{figure}

Parasitic phenomena, such as Fabry-Pérot interferences, were filtered using a two-step process implemented in MATLAB. First, a low-pass filter was applied to the raw reflectivity data to isolate slow variations caused by multiple reflections between parallel interfaces. This smoothing was performed using a moving average:

\[
y_s(i) = \frac{1}{N} \sum_{j=i-(N-1)/2}^{i+(N-1)/2} y(j),
\]

where \(N\) is the smoothing window size. The smoothed signal \(y_s(i)\) was then subtracted from the raw data \(y(i)\) to remove low-frequency oscillations:

\[
y_f(i) = y(i) - y_s(i).
\]
Parasitic phenomena, such as Fabry-Pérot interferences, were filtered using a two-step process implemented in MATLAB. First, a low-pass filter was applied to the raw reflectivity data to isolate slow variations caused by multiple reflections between parallel interfaces. This smoothing was performed using a moving average:
\[
y_s(i) = \frac{1}{N} \sum_{j=i-(N-1)/2}^{i+(N-1)/2} y(j),
\]
where \( N \) is the smoothing window size. The smoothed signal \( y_s(i) \) was then subtracted from the raw data \( y(i) \) to remove low-frequency oscillations:
\[
y_f(i) = y(i) - y_s(i).
\]
This filtered signal, denoted \( \Delta R \), corresponds to a differential reflectivity that emphasizes the high-frequency spectral variations associated with photonic modes, while effectively suppressing background oscillations and Fabry-Pérot interference (fig.\ref{S5}).

\newpage
\section{Resonance analysis and linewidth}
In resonance analysis, the Lorentzian function plays a crucial role in understanding the changes in intensity around the resonance point. A Lorentzian function \( I(E) \) is typically used to model a symmetric resonance:

\begin{equation}
    I(E) = \frac{I_0}{(E - E_0)^2 + \tau^2},
\end{equation}

where $I_0$ is the maximum intensity, $E_0$ is the resonance energy (central energy of the peak), and $\tau$ is the full-width at half maximum (FWHM) of the peak, related to the linewidth.

\begin{figure}[hbt!]
    \centering
\includegraphics[width=0.85\linewidth]{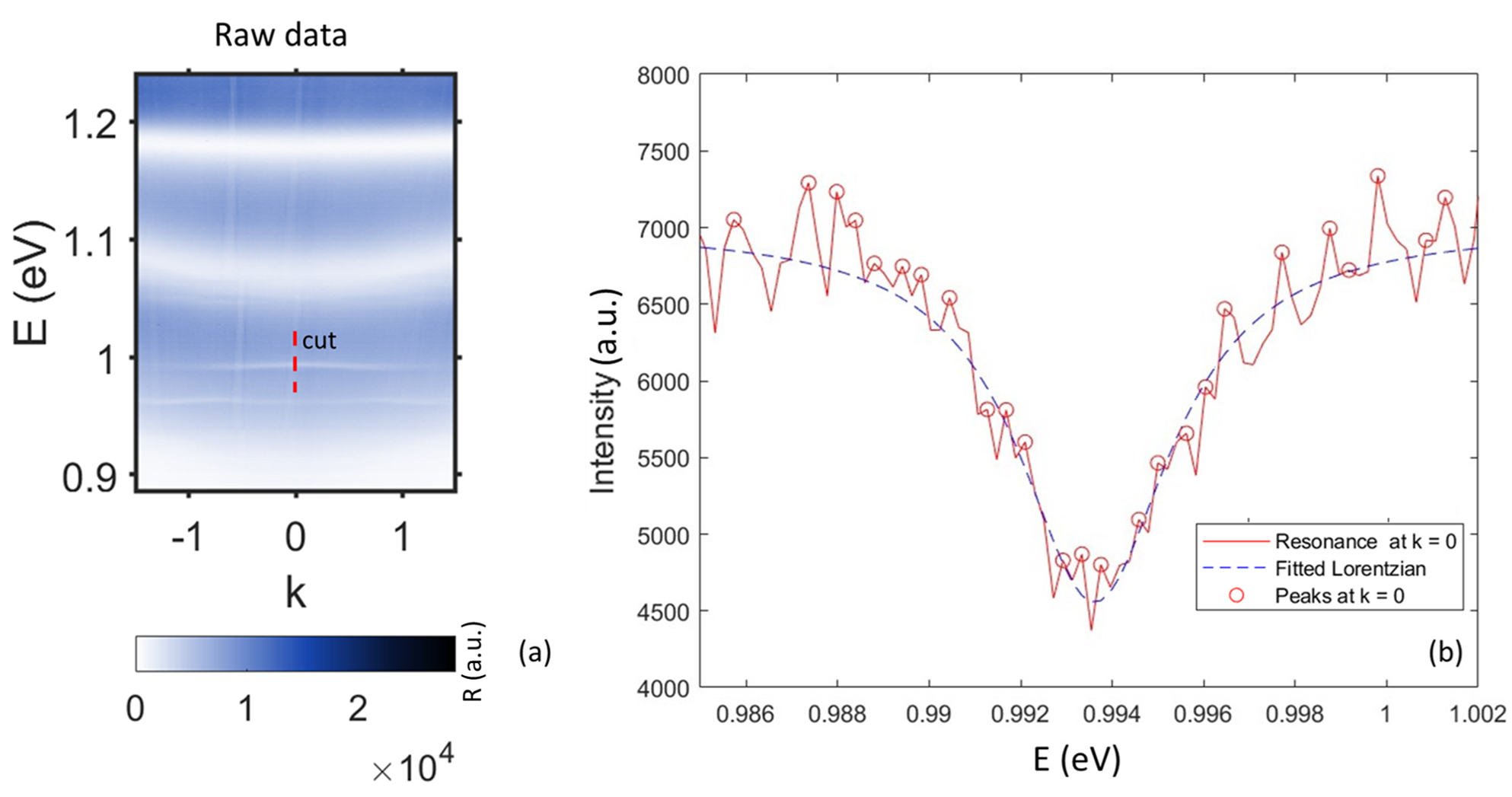}
    \caption{(a) Band diagram for the structure, with the studied cut region highlighted in red at \( k = 0 \). 
(b) Intensity profile as a function of energy \( E \) at \( k = 0 \), where the dashed blue curve represents a Lorentzian fit of the experimental data, and the red circles indicate the detected experimental peaks.
}
    \label{S6}
\end{figure}

Here, we use of the Lorentzian function to model the symmetric resonance observed in the experimental data. The Lorentzian function describes how the intensity \( I(E) \) varies around the resonance energy \( E_0 \), with the peak intensity at \( E_0 \) and a characteristic width determined by the parameter \( \tau \), which is related to the linewidth or full width at half maximum (FWHM) of the resonance.  

Figure \ref{S6}.a shows raw data as a color-coded plot of energy versus momentum, with the red dashed line indicating the specific data cut at \( k = 0 \), where further analysis is performed. We compare the experimental intensity data (in red) to the fitted Lorentzian curve (in blue), showing how the resonance behavior is captured by the model (fig.\ref{S6}.b). 

The fit allows the extraction of key parameters, such as the resonance energy \( E_0 \) (around 0.993 eV) and the linewidth (FWHM), which is calculated as \( 2 \tau \). In this case, if \( \tau \) is approximately 0.002 eV, the FWHM would be 0.004 eV (or 4 meV). Additionally, the quality factor ($Q$), which measures the sharpness of the resonance, can be calculated as the ratio of \( E_0 \) to the FWHM. 
For instance, with \( E_0 = 0.993 \, \text{eV} \) and FWHM = 0.004 eV, the quality factor \( Q \) would be approximately 250.

\end{document}